\documentclass[nofootinbib,prl,superscriptaddress,a4paper,twocolumn]{revtex4-1}
\usepackage{geometry}
\usepackage[english]{babel}
\usepackage[latin1]{inputenc}
\usepackage{hyperref}
\usepackage{amsmath, amssymb, amsfonts, mathrsfs}
\usepackage{graphicx}

\usepackage{mathtools}

\usepackage{algpseudocode}
\usepackage{algorithm}
\algrenewcommand\algorithmicrequire{\textbf{Assume:}}

\usepackage{braket}
\usepackage{float} 
\usepackage{silence}
\usepackage{url}
\usepackage{tabulary}
\usepackage{amsfonts}

\newcommand*{\melvin}{{\small M}{\scriptsize ELVIN}}

\newcounter{algostep}
\newlength{\stepwidth}

\graphicspath{{images/}}
\usepackage{xcolor}
\usepackage{exscale}
\usepackage{latexsym}
\usepackage{bbm}
\usepackage{xspace}

\newcommand{\beginsupplement}{%
        \setcounter{table}{0}
        \renewcommand{\thetable}{S\arabic{table}}%
        \setcounter{figure}{0}
        \renewcommand{\thefigure}{S\arabic{figure}}%
     }

\def\pmx{\begin{pmatrix}}
\def\emx{\end{pmatrix}}

\newcommand{\floors}[1]{ \left\lfloor  #1 \right\rfloor}

\definecolor{darkgreen}{RGB}{0,130,0}

\begin{document} 

\title{Arbitrary $d$-dimensional Pauli $X$-Gates of a flying Qudit}

\author{Xiaoqin Gao}
\email{xiaoqin.gao@univie.ac.at}
\affiliation{Vienna Center for Quantum Science \& Technology (VCQ), Faculty of Physics, University of Vienna, Boltzmanngasse 5, 1090 Vienna, Austria.}
\affiliation{Institute for Quantum Optics and Quantum Information (IQOQI), Austrian Academy of Sciences, Boltzmanngasse 3, 1090 Vienna, Austria.}
\affiliation{National Mobile Communications Research Laboratory, Southeast University, Sipailou 2, 210096 Nanjing, China.}
\affiliation{Quantum Information Research Center of Southeast University, Southeast University, Sipailou 2, 210096 Nanjing, China.}
\author{Mario Krenn}
\email{mario.krenn@univie.ac.at}
\affiliation{Vienna Center for Quantum Science \& Technology (VCQ), Faculty of Physics, University of Vienna, Boltzmanngasse 5, 1090 Vienna, Austria.}
\affiliation{Institute for Quantum Optics and Quantum Information (IQOQI), Austrian Academy of Sciences, Boltzmanngasse 3, 1090 Vienna, Austria.}
\author{Jaroslav Kysela}
\affiliation{Vienna Center for Quantum Science \& Technology (VCQ), Faculty of Physics, University of Vienna, Boltzmanngasse 5, 1090 Vienna, Austria.}
\affiliation{Institute for Quantum Optics and Quantum Information (IQOQI), Austrian Academy of Sciences, Boltzmanngasse 3, 1090 Vienna, Austria.}
\author{Anton Zeilinger}
\email{anton.zeilinger@univie.ac.at}
\affiliation{Vienna Center for Quantum Science \& Technology (VCQ), Faculty of Physics, University of Vienna, Boltzmanngasse 5, 1090 Vienna, Austria.}
\affiliation{Institute for Quantum Optics and Quantum Information (IQOQI), Austrian Academy of Sciences, Boltzmanngasse 3, 1090 Vienna, Austria.}

\begin{abstract}
High-dimensional degrees of freedom of photons can encode more quantum information than their two-dimensional counterparts. While the increased information capacity has advantages in quantum applications (such as quantum communication), controlling and manipulating these systems has been challenging. Here we show a method to perform lossless arbitrary high-dimensional Pauli-X gates for single photon. The $X$-gate consists of a cyclic permutation of qudit basis vectors, and, together with the $Z$ gate, forms the basis for performing arbitrary transformations. We propose an implementation of such gates on the orbital angular momentum of photons. The proposed experimental setups only use two basic optical elements such as mode-sorters and mode-shifters -- thus could be implemented in any system where these experimental tools are available. Furthermore the number of involved interferometers scales logarithmically with the dimension, which is important for practical implementation.

\end{abstract}

\date{\today}
\maketitle

\section{Introduction}

High-dimensional quantum systems allow for encoding, transmitting and processing more than one bit per photon. The exploitation of large alphabets in quantum communication protocols can significantly improve their capacity \cite{S2006experimental, mirhosseini2015high, sit2017high} and security \cite{huber2013weak, bouchard2017high}. However, performing well-defined manipulations in multi-level systems is significantly more challenging than for qubits. Schemes for implementation of an arbitrary unitary transformation were developed for the path degree of freedom already in 1994 \cite {Zeilinger1994, carolan2015universal, schaeff2015experimental}. However, path encoding schemes are very susceptible to phase changes and it is thus very challenging to use them in real-world quantum communication. Laguerre-Gaussian modes of light carrying orbital angular momentum (OAM) \cite {allen1992orbital,  padgett2017orbital} represent an alternative to polarization that has become a popular choice in experiments in a high-dimensional quantum domain \cite {mair2001entanglement, rubinsztein2016roadmap, Manuel2018}. In contrast, the OAM of photons has been used successfully in long-distance classical \cite {krenn2014communication,  krenn2016twisted, ren2016experimental, lavery2017free} as well as quantum \cite {krenn2015twisted, sit2017high, bouchard1801underwater} communication in the form of a ''flying qudit''. An important open question is how an arbitrary high-dimensional transformation can be performed with OAM.

An arbitrary unitary transformation in a finite-dimensional space can be expressed as a combination of Pauli $X$- and $Z$-gates and their integer powers \cite {Asadian2016}, which shows that they are important basis gates. The $Z$ gate for OAM qudit introduces a mode-dependent phase, which can be implemented simply by a Dove prism \cite {Leach2002, gonzalez2006dove, agnew2013generation, zhang2016engineering}. The $X$-gate in high-dimensional Hilbert spaces takes the form of a cyclic permutation of the computational basis vectors. For a fixed basis in a $d$-dimensional space the cyclic transformation transforms each basis state into its nearest neighbor in a clockwise manner with the last state being transformed back to the first one \cite {Schlederer2016}. As such, the cyclic transformation is the $d$th root of unity that performs a rotation in a $d$-dimensional space. While efficient methods for realizing a four-dimensional cyclic transformation both in classical and quantum realms have been experimentally demonstrated \cite {Schlederer2016, Babazadeh2017, Wang2017}, for arbitrary dimension such methods are still missing.

Here we present setups for arbitrary $d$-dimensional $X$-gates represented by the OAM of single photons. When developing the general solution we took inspiration in designs generated by the computer program \melvin \  \cite {Krenn2016}. Schemes produced by the solution can be implemented in the laboratory using accessible optical components. The setups employ only two basic elements: holograms and OAM beam-splitters (OAM-BSs) introduced by Leach \emph{et al.} \cite {Leach2002}.Importantly, the number of OAM-BSs (which are interferometric devices) scales logarithmically with the dimension of the cycle, which is relevant for their experimental implementation.

\begin{figure*}
  \centering
  \begin{minipage}[t]{0.86\textwidth}
    \centering
    \raisebox{-\height}{\includegraphics[width=1\textwidth]{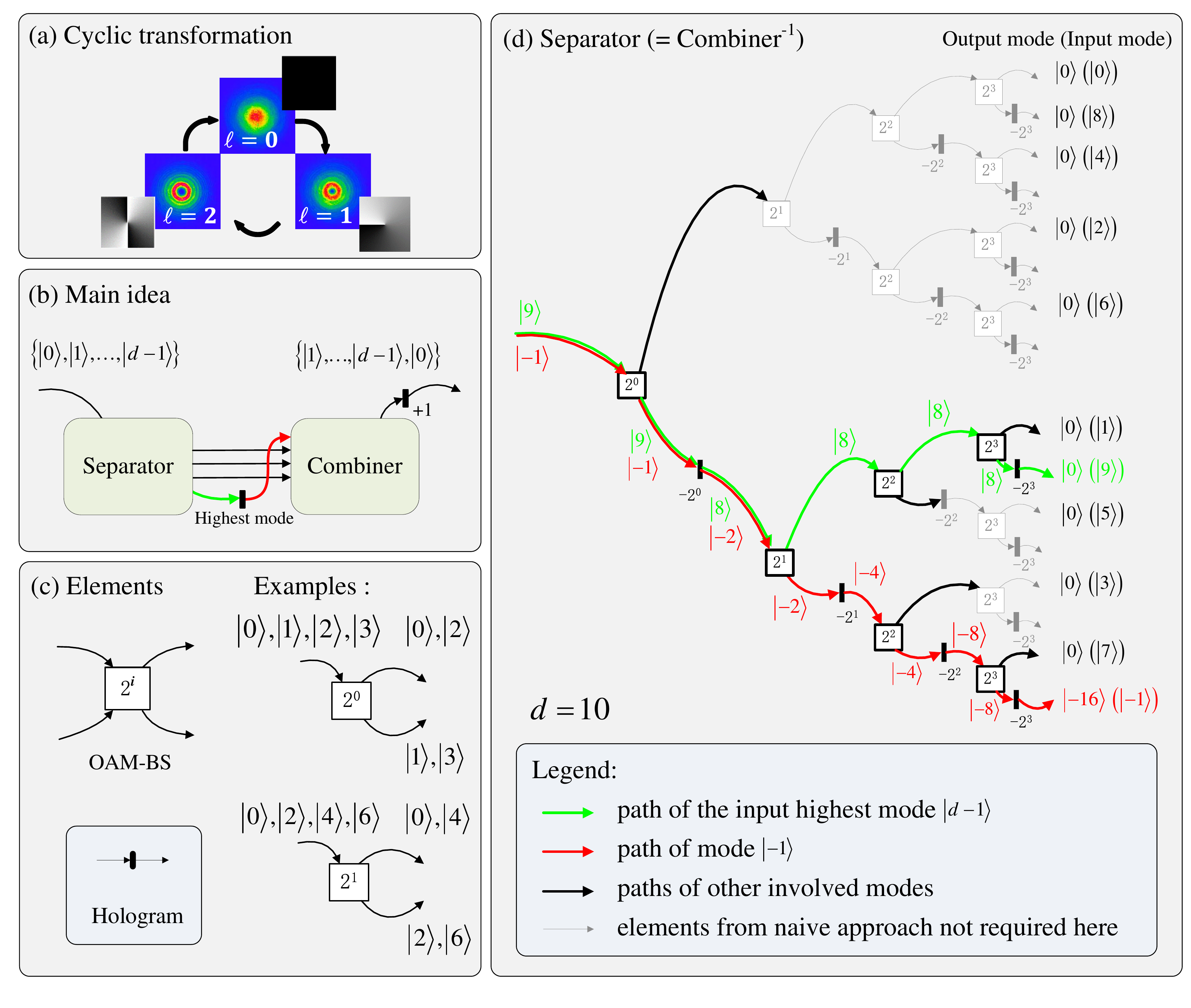}}
    \vspace{-0.4cm}%
    \caption{Working principle of the general solution. (a) The cyclic transformation in a three-dimensional space spanned by OAM modes $\ell$=0, 1, and 2. If the incoming beam possesses OAM of $\ell=0$ ($\ell=1$), it becomes $\ell=1$ ($\ell=2$) after adding +1 to the mode. The highest OAM mode $\ell=2$ is transformed into $\ell=0$ by subtraction of 2 quanta of OAM. The experimentally obtained intensity profiles and computer-generated phase profiles of OAM modes in a beam cross-section are inserted for convenience. (b) Main idea: The general setup is combined by a Separator, a Combiner, a +1 mode shift and an additional hologram that is inserted in the path of the highest mode.  The Separator splits the mode $\ket{d-1}$ (green) and $\ket{-1}$ (red; it is an ancillary mode that is required in our scheme) into separate paths. Then the OAM of the highest mode is changed to $\ket{-1}$, and it enters the path of mode $\ket{-1}$. Subsequently, all modes are combined again by the Combiner into one path and acquire +1 quantum of OAM in the end. (c) Two basic elements:  OAM-BSs and holograms. The label $2^i$ determines the sorting properties of the OAM-BSs. Holograms perform shifts of the OAM value by a fixed predefined amount. See Supplemental Material  \cite {XQ} for details about the structures of the OAM-BSs and holograms. (d) Separator: we choose the highest mode  $\ket{9}$ (green) in the 10-dimensional cycle. Afterwards, it will enter into the lowest path of mode $\ket{-1}$. The Separator and Combiner display a high degree of symmetry --- they are mirror reflections of each other, with an inversion of hologram values.  It is instructive to compare our scaling behavior with the naive approach, where $(d-1)$ OAM-BSs are used to reroute all $d$ OAM modes into separate paths, $d$ holograms are then used to shift OAM values of all modes, and finally additional $(d-1)$ OAM-BSs are necessary to recombine all modes together. Our approach therefore offers an exponentially smaller number of interferometers than the naive approach.}
\label{fig1:figure1}
\vspace{-0.5cm}%  
\end{minipage}\hfill
\end{figure*}

\section{Arbitrary $d$-dimensional $X$-gates}

The main goal of this paper is to find lossless experimental setups for realizing arbitrary $d$-dimensional $X$-gates with the OAM of single photons. We would like to perform an $X$-gate that consists of a cyclic permutation of qudit basis vectors. One example is the cyclic transformation in three-dimensional space shown in figure \ref{fig1:figure1}(a), where the OAM mode $\ell=0$  is transformed into $\ell=1$, the mode $\ell=1$ into $\ell=2$, and $\ell=2$ back into $\ell=0$. The main idea behind our approach is shown in figure \ref{fig1:figure1}(b): The setup consists of a Separator, a Combiner, a +1 mode shifter, and an additional mode shifter that works only for the highest mode. The  Separator is used for splitting modes from one path to multiple paths, while the Combiner does the opposite, regrouping all modes into a single path. First, the input photon that is in the subspace spanned by the modes  $\ket{0},\ket{1},\ket{2},\ldots,\ket{d-1}$, goes through the Separator, which sends the highest mode to its own path. Subsequently, the highest mode acquires a mode shift and  enters the path of mode $\ket{-1}$. Finally, the Combiner sends all modes into one single path and then a mode shifter adds +1 quantum of OAM. The Separator and the Combiner consist of multiple OAM-BSs and holograms, respectively, as shown in figure \ref{fig1:figure1}(c). The Separator and the Combiner display a high degree of symmetry: The Combiner is a mirror reflection of the Separator, with an inversion of the hologram values. Except for the highest mode, all other modes only add +1 quantum of OAM in total. In our method, we only need to separate the path of mode $\ket{-1}$ and the path of the highest mode from all other modes. As an example, we show the ten-dimensional cycle in figure \ref{fig1:figure1}(d). The highest mode $\ket{9}$ will enter the lowest path of mode $\ket{-1}$ (red) after going through a mode shift of -16 quantum of OAM. Due to the symmetry of the Separator and the Combiner, mode $\ket{9}$ will change to mode $\ket{-1}$ before the last mode shifter adds +1. Therefore, the highest mode ends up in the mode $\ket{0}$, as required. 
%%\begin{align}
%d-1= \sum_{i=0}^{N-1}(b_i\times2^i), \quad b_i=0,1.
%\label{eq:equation1}
%\end{align}

\section{Scaling}
The proposed experiments for $X$-gates involve interferometers, which are demanding to stabilize experimentally. For that reason, a small number of interferometers is favorable. One naive approach to perform a $X$-gate would be to transform the OAM information to path encoding by splitting every mode into its own path. This would require $2(d-1)$ interferometers, which is linear in the dimension. The strength of our method is, that it requires only a logarithmical number of interferometers, thus is significantly easier to implement, especially for high dimensions.

Based on the main idea of figure \ref{fig1:figure1}, one can draw the setup with arbitrary dimension $d$. In general, the setup for the $d$-dimensional cycle is a combination of two different experimental structures, namely one for powers-of-two $2^M$ and one for odd numbers $Q$. The total dimension is given by 
\begin{align}
d = 2^M\times Q.
\label{eq:equation2}
\end{align}

For example, the setup for the case $d=88=8 \times 11$ is built up from two setups of dimension $d=8$ and $d=11$ (See Supplemental Material  \cite {XQ}).

Remarkably, the number $N_{Arb}$ of OAM-BSs scales logarithmically with the dimension $d$, as follows \footnote{ These numbers are obtained with a slightly optimized setup, where for cycles with $Q>1$ two unnecessary OAM-BSs and unnecessary holograms are removed. An algorithmic code to produce cyclic transformation setups for arbitrary $d$ is shown in  Supplemental Material  \cite {XQ}.}
\begin{equation}
\begin{aligned}
N_{Arb}(d) ={} &  N_{2^M}(d) + N_{Odd}(d)  \\
     & = 2 \times (M(d) + 2 \times \floors { \log_2 {Q(d)} }) \\
      & \leq 4 \times   { \log_2 {(d-1)} },
\label{eq:equation3}
\end{aligned}
\end{equation}
where $M(d)$ and $Q(d)$ can be represented using elementary operations \footnote{The power-of-two and the odd part of an integer $d$ can be obtained using these elementary functions: $M(d)=\sum_{n=1}^{d}( \lfloor \cos^{2}(\frac{d \* \pi} {2^n} ) \rfloor )$, $Q(d)=\frac{d} {2^{M(d)}} $.}.

% \footnote{$ M(d)=\sum_{n=1}^{d}( \floos {\cos^{2}(\frac{d \* \pi} {2^n})})$, $Q(d)=\frac{d} {2^{M(d)}} $}.

\begin{figure}[!htbp]
\includegraphics[width=\columnwidth]{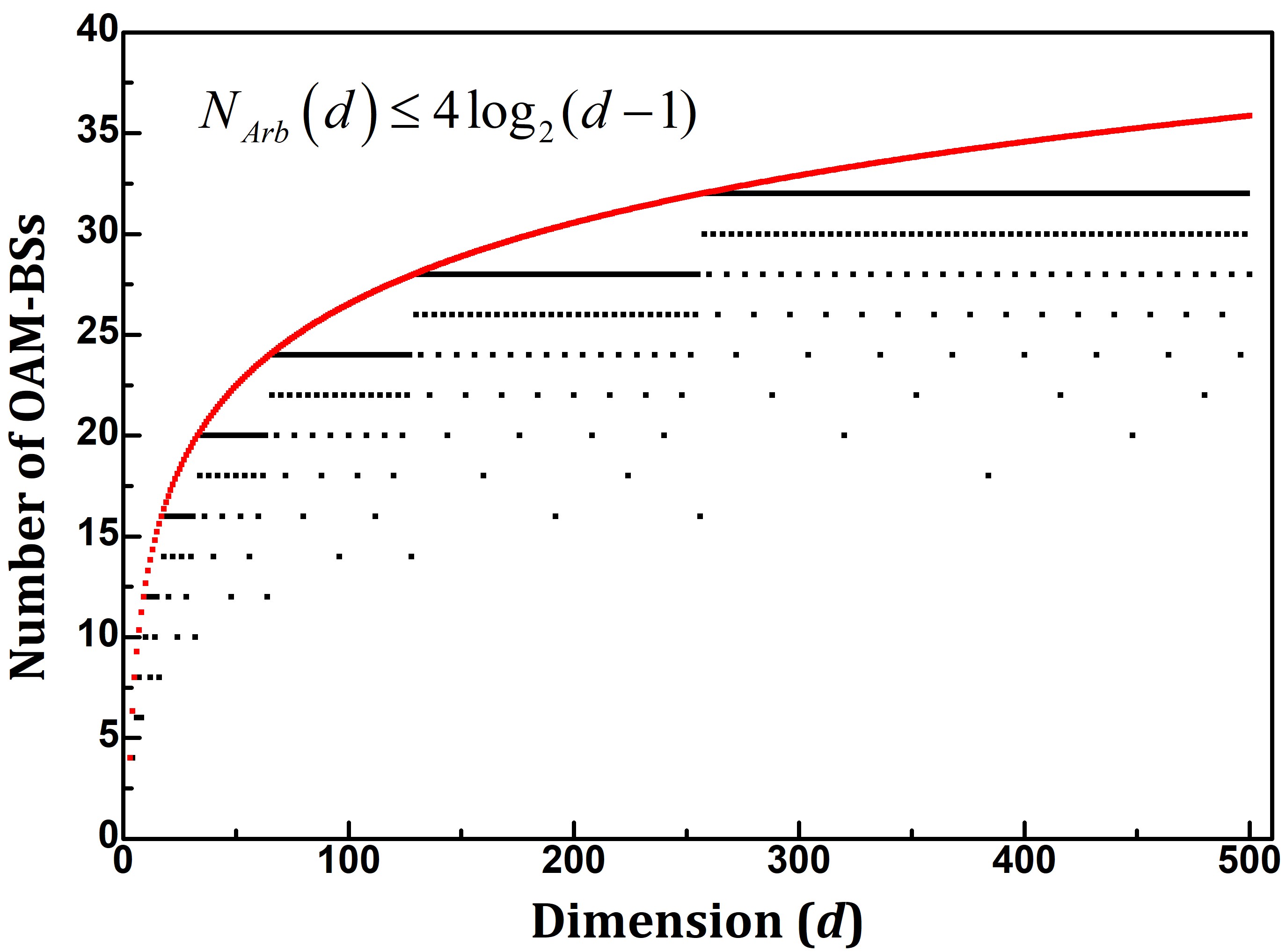}
 \vspace{-0.4cm}%
\caption{The number of OAM-BSs scales logarithmically with the dimension $d$.}  
\centering
\label{fig6:figure6}
\vspace{-0.1cm}%  
\end{figure}

This is shown in figure \ref{fig6:figure6}, where the number $N_{Arb}$ is plotted as a function of dimension $d$ for $3 \leq d \leq 500$. For example, we can realize 500-dimensional cyclic transformation using a setup with only 28 OAM-BSs. Such a setup is already within reach of the present-day technology --- recently an experiment has been reported \cite {wang201818} where 30 interferometers were kept stable over 72 hours without any active stabilization.

Interestingly, for any given $M \geq 1$ and for all odd dimensions $d$ in the range $2^M + 1, \ldots, 2^{M+1} - 1$ the number of OAM-BSs in the generated setups stays constant, as shown in figure \ref{fig6:figure6}, and only the connections between individual elements differ. For example, the number of OAM-BSs is 12 for all $d = 9, 11, 13$, and 15, as shown in  Supplemental Material  \cite {XQ}.

\begin{figure}[!htbp]
\includegraphics[width=\columnwidth]{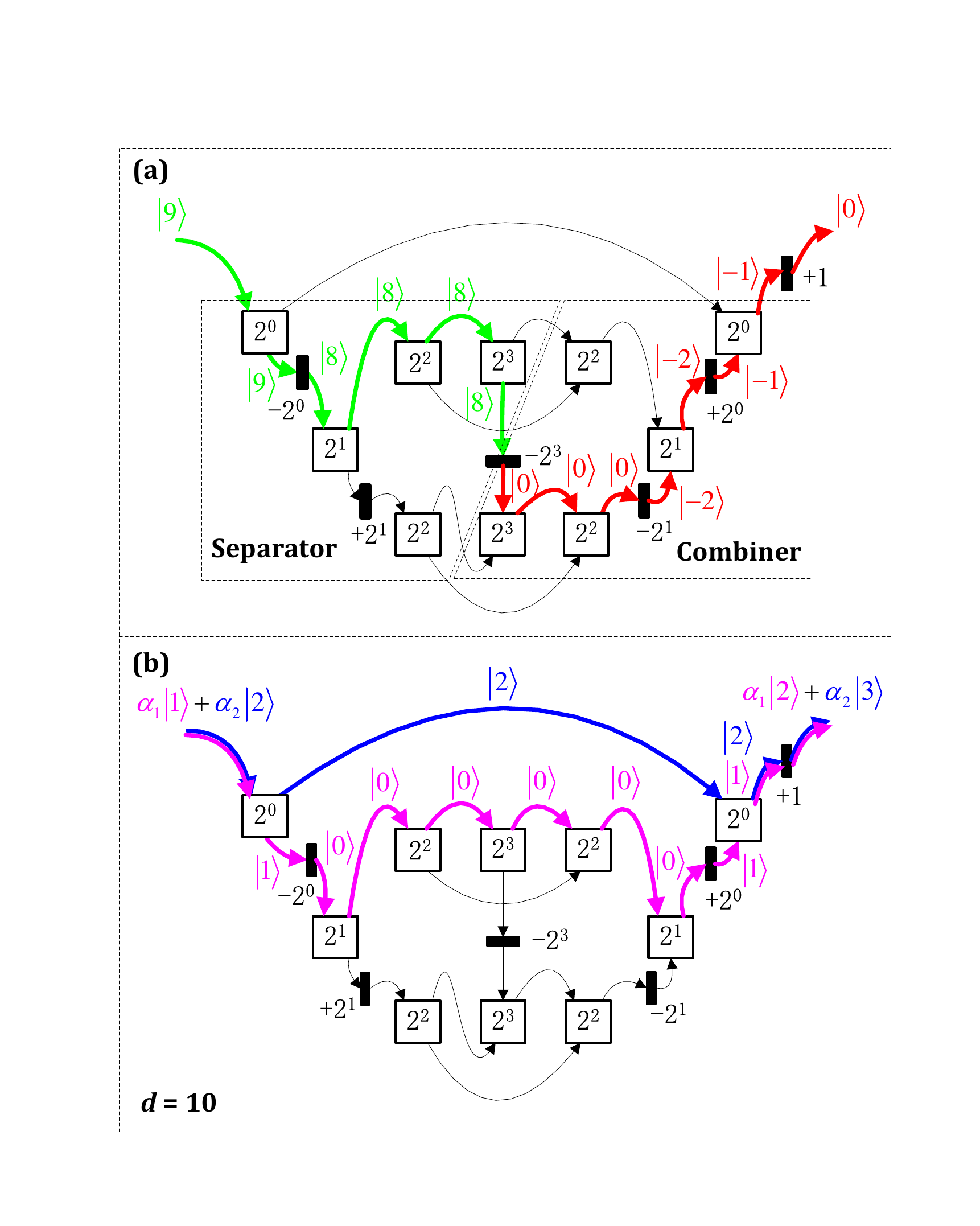}
 \vspace{-0.4cm}%
\caption{(a) The experimental setup for the $X$-gate in the 10-dimensional space. The Separator and the Combiner here correspond to the main idea in figure \ref{fig1:figure1}. Four different types of OAM-BSs and six holograms are used in the setup. Unlike other modes the highest-order mode $\ket{9}$ propagates through the middle path and enters the path of mode $\ket{-1}$ after receiving a -$2^3$ mode shift. (b) The setup also works for a superposition, such as $\ket{ \psi_{in} } =\alpha_1 \ket{1} +\alpha_2 \ket{2}$.  }  
\centering
\label{fig13:figure13}
\vspace{-0.1cm}%  
\end{figure}

As an illustration, we depict the setup for the case $d=10$ in figure \ref{fig13:figure13}. The Separator and the Combiner in figure \ref{fig13:figure13}(a) correspond to the main idea in figure \ref{fig1:figure1}. The propagation of three modes is shown explicitly in the figure in order to illustrate how the experimental implementation works. The crucial property of the setup is that only the highest-order mode  propagates through the center, hence it undergoes an additional $-2^3$ OAM shift and then enters the lowest path of mode $\ket{-1} $  resulting in the transformation $\ket{9} \to \ket{0}$ after the last operation +1. If the path differences are matched such that they are smaller than the coherence length of the input, the setup also works for a superposition.

\section{Conclusion and Outlook}

We developed a solution for experimental setups realizing arbitrarily high-dimensional cyclic transformations with OAM of single photons. The total number of required interferometers scales logarithmically with the dimension. Given that a recent experiment has been demonstrated \cite {wang201818} with very high quality and stability of 30 interferometers, our method is experimentally feasible in high dimensions. A cyclic transformation in a given $d$-di\-men\-sion\-al space is defined for a specific basis $\ket{0},\ket{1},\ket{2},\ldots,\ket{d-1}$,  but also for other sets of states without adaptation of the experimental setup,  which are shown in the Supplemental Material  \cite {XQ}.  Furthermore, we can also get arbitrary $d$-dimensional cyclic transformations for OAM modes $\ket{0+m},\ket{1+m},\ket{2+m},\cdots,\ket{d-1+m}$ if we put one hologram with value $m$ before the setup and another hologram with the opposite value $-m$ after the setup.  The structure of  the setups generated by our method is very symmetric. Consequently, if the photon has a sufficiently large coherence length, the original setup can be considerably simplified. This is shown in the  Supplemental Material  \cite {XQ}. Implementations of $X$-gates generated by the method presented here have a very convenient property that they can be converted into the $X^{-1}$ gate implementations only by a modification of two holograms and slight reconnection of two OAM-BSs, as shown in the   Supplemental Material  \cite {XQ}. This is beneficial in future implementations, where quick automated changes between gates are necessary. There are still some very interesting emerging questions. 

To experimentally create arbitrary unitary transformations one needs to combine all of the $X^l Z^m$ building blocks. This recombination can be performed in the probabilistic way, which leads to losses. An important immediate question is how our results can be generalized so that broader classes of transformations can be realized in a lossless deterministic manner.

The high-dimensional generalized controlled-NOT (CNOT) gate is a controlled cyclic gate. Realizing arbitrary $d$-dimensional CNOT gates in the OAM would be very desirable, as it allows for more complex control and processing of multi-photonic states.

The method works in principle in any quantum system where experimental tools for mode sorting and mode shifting exist. In particular, the full transverse structure of photons consists of the orbital angular momentum mode (investigated here) and the radial mode ($p$-mode) \cite {karimi2014radial, plick2015physical}. Lossless mode sorters for radial modes have recently been implemented  \cite {zhou2017sorting, gu2018gouy}. Thus, in order to generalize our result to radial modes, one requires a method to increase the mode number of radial modes -- which is an interesting path for future research.

Our scheme for implementing arbitrary dimensional $X$-gates was discovered by interpreting and generalizing special-case solutions found by the computer program \melvin \cite{Krenn2016}. Various other computer programs have recently been developed to autonomously find, optimize or simplify quantum experiments \cite{knott2016search, melnikov2018active, arrazola2018machine}. It is exciting to think about how a computer program itself would be able to generalize special-case solutions in a similar way as human intelligence.

The authors thank Giuseppe Vitagliano, Nicolai Friis, Jessica Bavaresco and Maximilian Lock for helpful discussions. XQ. G. thanks Bin Sheng and Zaichen Zhang for support. This work was supported by the Austrian Academy of Sciences ({\"O}AW), and the Austrian Science Fund (FWF) with SFB F40 (FOQUS). XQ. G. acknowledges support from the National Natural Science Foundation of China (No. 61571105), the Hong Kong, Macao and Taiwan Science \& Technology Cooperation Program of China (2016YFE0123100), the Scientific Research Foundation of Graduate School of Southeast University (No. YBJJ1710), and China Scholarship Council (CSC).

\bibliographystyle{unsrt}
\bibliography{refs}

\clearpage

\section{Supplementary}
\beginsupplement

\subsection{OAM-BSs and holograms}
\begin{figure}[!htbp]
\includegraphics[width=0.48 \textwidth]{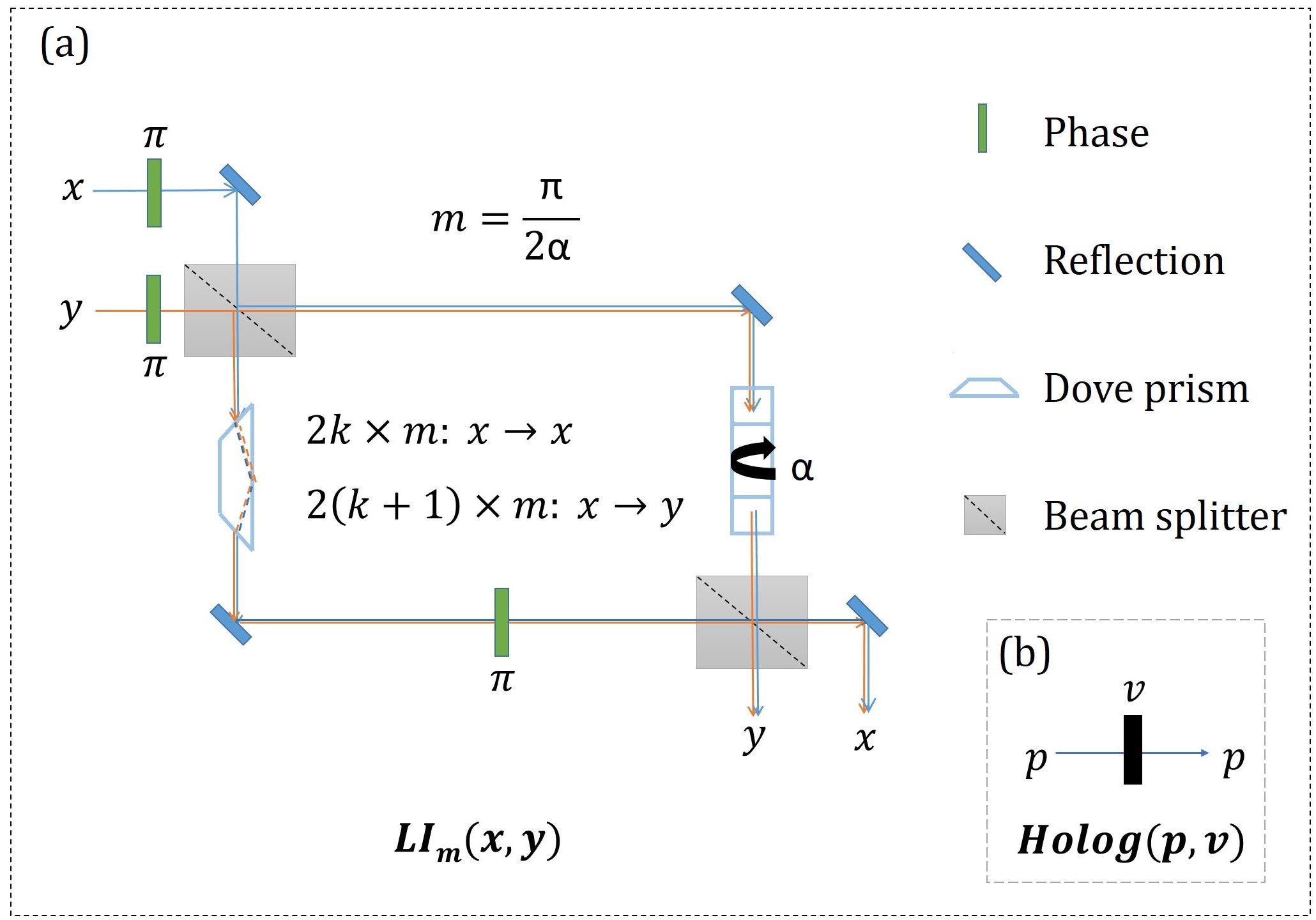}
\caption{(a) An OAM beam-splitter, denoted by $LI_m(x, y)$, where $x$ and $y$ are input and output paths and $m$ specifies the angle of one Dove prism, $m=\pi / {2 \alpha}$.  If the input mode is $2k \times m$ ($k \in \mathbb{Z}$), the output path is the same as the input path. For $(2k+1) \times m$ it will go to the other path. For $m=1$ the OAM-BS works as a parity sorter. (b) A hologram, denoted by $Holog(p, v)$, where $p$ is the path and $v$ stands for the value added to the mode.}  
\centering
\label{fig2:figure2}
\vspace{-0.5cm}%  
\end{figure}

The structure of the OAM-BS is depicted in figure \ref{fig2:figure2}(a) --- it consists of a Mach-Zehnder interferometer with a Dove prism in each arm. The angle $\alpha$, by which one Dove prism is rotated with respect to the other, determines the sorting properties of the OAM-BS. If the OAM value of the incoming photon is an even multiple of $m=\pi / {2 \alpha}$, the photon leaves the OAM-BS from one port. If the OAM value is an odd multiple of $m$, it exits from the other port. We can therefore stack multiple OAM-BSs with successively increasing $m$ to separate each mode by routing it into a different path. For $m = 1$ the OAM-BS works as a parity sorter. We denote the OAM-BS by $LI_m(x,y)$ in the following, where $x$ and $y$ are input and output paths and $m$ is as discussed above. Holograms perform shifts of the OAM value by a fixed predefined amount, as shown in figure \ref{fig2:figure2}(b). By $Holog(p,v)$ we designate a hologram that adds $v$ quanta of the OAM to a mode in path $p$.

%\vspace{-1cm}

\newpage
\subsection{Pseudocodes}

\begin{minipage}{0.9\linewidth}
\begin{algorithm} [H]
   \caption{Pseudocode for Arbitrary Dimension}
  \label{alg:algorithm3}
  \begin{algorithmic}%[3]
 \Require{The input state is a linear combination of modes $\ket{0}, \ket{1}, \ket{2}, \cdots, \ket{d-1}$ and enters the first path $r_0$.}
 \Require{$LI_m(x, y)$ stands for an OAM-BS, where $m$ determines into which paths ($x$ or $y$) individual OAM modes are output, as shown in figure \ref{fig2:figure2}(a).}
 \Require{$Holog(p, v)$ stands for a hologram in path $p$ that adds value $v$ to the OAM mode. It is shown in figure \ref{fig2:figure2}(b).}
\Require{$N$ and $b_t$ are calculated from $Q = \sum_{i=0}^{N-1}(b_i\times2^i), \quad b_i=0,1.$  $M$ is calculated by $d = 2^M\times Q$.}

\Require{Auxiliary indices $a_i$ are defined as:
\State  $ a_{1}$ = 0
\For{$t \textrm{ from } 1 \textrm{ to } N-2$}
         \If{$b_t=0$}
          \State $ a_{t+1}$ = $ a_{t}$ 
        \Else 
              \State $ a_{t+1}$ = $t$
        \EndIf
\EndFor}
\\

\Procedure{Cycle for Arbitrary Dimension}{}
 
\For{\textcolor{purple} {$t \textrm{ from } 0 \textrm{ to } M-1$} }
\State \textcolor{purple}{$LI_{2^t}(r_{t}, r_{t+1})$}
\State   \textcolor{purple}{$Holog(r_{t+1}, -2^{t})$}
\EndFor

 \If{$N=1$}
          \State  $Holog(r_{M}, -2^{M})$
\Else 

\State   $ LI_{2^{M}}(r_{M}, s_0)$
\State  $Holog(s_0, +2^{M})$

\For{\textcolor{blue} {$t \textrm{ from } 1 \textrm{ to } N-2$}}
          \State \textcolor{blue} {$LI_{2^{t+M}}(r_{a_{t}+M}, r_{t+M})$}
         \State \textcolor{blue}{$Holog(r_{t+M}, -{b_{t}}\times{2^{t+M}})$}
\EndFor

\State $ LI_{2^{N-1+M}}(r_{a_{N-1}+M}, r_{N-1+M})$
\State $ Holog(r_{N-1+M}, -2^{N-1+M})$

\For{\textcolor{blue} {$t \textrm{ from } N-2 \textrm{ down to } 1$}}
     \State\textcolor{blue} {$Holog(r_{t+M}, +{b_{t}}\times{2^{t+M}})$}
                  \State\textcolor{blue} {$LI_{2^{t+M}}(r_{a_{t}+M}, r_{t+M})$}
\EndFor

\For{\textcolor{darkgreen} {$t \textrm{ from } 1 \textrm{ to } N-2$}}
         \State  \textcolor{darkgreen}{$LI_{2^{t+M}}(s_0, s_t)$}
\EndFor

\State $LI_{2^{N-1+M}}(r_{N-1+M}, s_0)$

\For{\textcolor{darkgreen} {$t \textrm{ from } N-2 \textrm{ down to } 1$}}
        \State \textcolor{darkgreen}{$LI_{2^{t+M}}(r_{N-1+M}, s_{t})$}
 \EndFor
  
\State $Holog(r_{N-1+M}, -2^M)$
\State $LI_{2^M}(r_M, r_{N-1+M})$

\EndIf
     
\For{\textcolor{purple}{$t \textrm{ from } M-1 \textrm{ down to } 0$}}
      \State\textcolor{purple}{$Holog(r_{t+1}, +2^t)$}
     \State \textcolor{purple}{$LI_{2^{t}}(r_{t}, r_{t+1})$}
\EndFor

\State $Holog(r_0, +1)$

\EndProcedure
  \end{algorithmic}
\end{algorithm}
\end{minipage}

\begin{minipage}{0.9\linewidth}
\begin{algorithm} [H]
   \caption{Special case --- Pseudocode for Odd Dimension}
  \label{alg:algorithm1}
  \begin{algorithmic}%[1]
 \Require{The input state is a linear combination of modes $\ket{0}, \ket{1}, \ket{2}, \cdots, \ket{d-1}$ and enters the first path $r_0$.}
 \Require{$LI_m(x, y)$ stands for an OAM-BS, where $m$ determines into which paths ($x$ or $y$) individual OAM modes are output, as shown in figure \ref{fig2:figure2}(a).}
 \Require{$Holog(p, v)$ stands for a hologram in path $p$ that adds value $v$ to the OAM mode. It is shown in figure \ref{fig2:figure2}(b).}
 \Require{$N$ and $b_t$ are calculated via  $d = \sum_{i=0}^{N-1}(b_i\times2^i), \quad b_i=0,1.$ }

 \Require{Auxiliary indices $a_i$ are defined as:
\State  $ a_{1}$ = 0
\For{$t \textrm{ from } 1 \textrm{ to } N-2$}
         \If{$b_t=0$}
          \State $ a_{t+1}$ = $a_{t}$ 
        \Else 
              \State $ a_{t+1}$ = $t$
        \EndIf
\EndFor}

\\

\Procedure{Cycle for Odd Dimension}{}

\State $LI_{2^0}(r_0, s_0)$
\State $Holog(s_0, +1)$

\For{\textcolor{blue}{$t \textrm{ from } 1 \textrm{ to } N-2$}}
      \State \textcolor{blue} {$ LI_{2^{t}}(r_{a_{t}}, r_{t})$}
     \State \textcolor{blue} {$Holog(r_{t}, -{b_{t}}\times{2^{t}})$ }
\EndFor

\State $LI_{2^{N-1}}(r_{a_{N-1}}, r_{N-1})$
\State $Holog(r_{N-1}, -2^{N-1})$

\For{\textcolor{blue} {$t \textrm{ from } N-2 \textrm{ down to } 1$}}
     \State\textcolor{blue}{$Holog(r_{t}, +{b_{t}}\times{2^{t}})$}
     \State \textcolor{blue}{$LI_{2^{t}}(r_{a_{t}}, r_{t})$}
\EndFor
  
\For{\textcolor{darkgreen} {$t \textrm{ from } 1 \textrm{ to } N-2$}}
         \State  \textcolor{darkgreen}{$LI_{2^{t}}(s_0, s_t)$}
\EndFor

\State  $LI_{2^{N-1}}(r_{N-1}, s_0)$

\For{\textcolor{darkgreen}{$t \textrm{ from } N-2 \textrm{ down to } 1$}}
         \State \textcolor{darkgreen}{$LI_{2^{t}}(r_{N-1}, s_t)$}
\EndFor
  
\State $Holog(r_{N-1}, -1)$
\State $LI_{2^0}(r_0, r_{N-1})$
\State $Holog(r_0, +1)$

\EndProcedure

\end{algorithmic}
\end{algorithm}
\end{minipage}
\\

\begin{minipage}{0.9\linewidth}
\begin{algorithm} [H]
   \caption{Special case --- Pseudocode for Power-of-Two Dimension}
\centering
  \label{alg:algorithm2}
  \begin{algorithmic}%[2]
 \Require{The input state is a linear combination of modes $\ket{0}, \ket{1}, \ket{2}, \cdots, \ket{d-1}$ and enters into path $r_0$.}
 \Require{$LI_m(x, y)$ stands for an OAM-BS, where $m$ determines into which paths ($x$ or $y$) individual OAM modes are output, as shown in figure \ref{fig2:figure2}(a).}
 \Require{$Holog(p, v)$ stands for a hologram in path $p$ that adds value $v$ to the OAM mode. It is shown in figure \ref{fig2:figure2}(b).}
\Require{$M = \log_2 d$.}

 \\

\Procedure{Cycle for $2^M$ Dimension}{}

\For{\textcolor{purple}{$t \textrm{ from } 0 \textrm{ to } M-1$}}
    
 \State\textcolor{purple}{$LI_{2^{t}}(r_{t}, r_{t+1})$}
  \State \textcolor{purple}{$Holog(r_{t+1}, -2^{t})$ }  
\EndFor

\State  $Holog(r_{M}, -2^{M})$

\For{\textcolor{purple}{$t \textrm{ from } M-1 \textrm{ down to } 0 $}}
  
   \State \textcolor{purple}{$Holog(r_{t+1}, +2^{t})$}
 \State \textcolor{purple}{$LI_{2^{t}}(r_{t}, r_{t+1})$}

\EndFor

\State $Holog(r_0, +1)$

\EndProcedure
%\centering
\end{algorithmic}
\end{algorithm}
\end{minipage}

\begin{figure*}
  \centering
  \begin{minipage}[t]{0.9\textwidth}
    \centering
    \raisebox{-\height}{\includegraphics[width=1\textwidth]{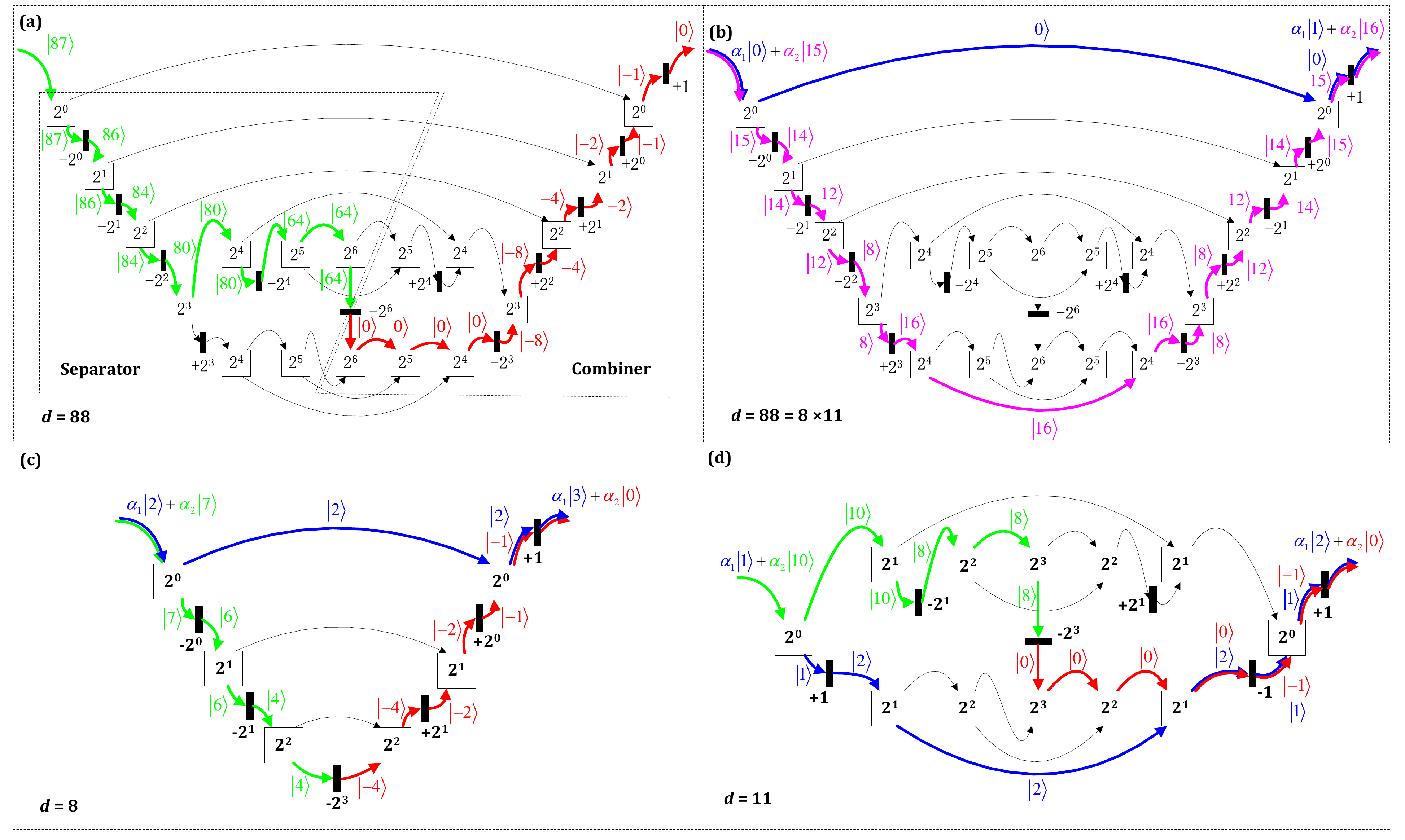}}
    \vspace{-0.3cm}%
    \caption{(a) The experimental setup for the $X$-gate in the 88-dimensional space. That is $d=88$, for which $M=3$ and $Q=11$ according to  $d = 2^M\times Q$. The setup is a combination of setups for $d = 8$ and $d = 11$.   Unlike other modes the highest-order mode $\ket{87}$ propagates through the middle path and enters the path of mode $\ket{-1}$. There are only 18 OAM-BSs in the setup for $d = 88$. (b) The setup works for a superposition, such as $\ket{ \psi_{in} } =\alpha_1 \ket{0} + \alpha_2 \ket{15}$. (c) The experimental setup for realization of the $X$-gate in an eight-dimensional space. The setup works for arbitrary complex linear combination of modes $\ket{0}$ through $\ket{7}$. For example, an input state $\ket{ \psi_{in}} = \alpha_1 \ket{2} + \alpha_2 \ket{7}$ is transformed into an output state $\ket {\psi_{out}} =  \alpha_1 \ket{3} + \alpha_2 \ket{0}$. (d) The experimental setup for the $X$-gate in dimension $d = 11$.  The device works for coherent superposition, as is shown for the example $\ket{ \psi_{in} } = \alpha_1 \ket{1} + \alpha_2 \ket{10}$. }
\label{fig5:figure5}
\vspace{-0.3cm}%  
\end{minipage}\hfill
\end{figure*}

\begin{figure*}
  \centering
  \begin{minipage}[t]{0.9\textwidth}
    \centering
    \raisebox{-\height}{\includegraphics[width=1\textwidth]{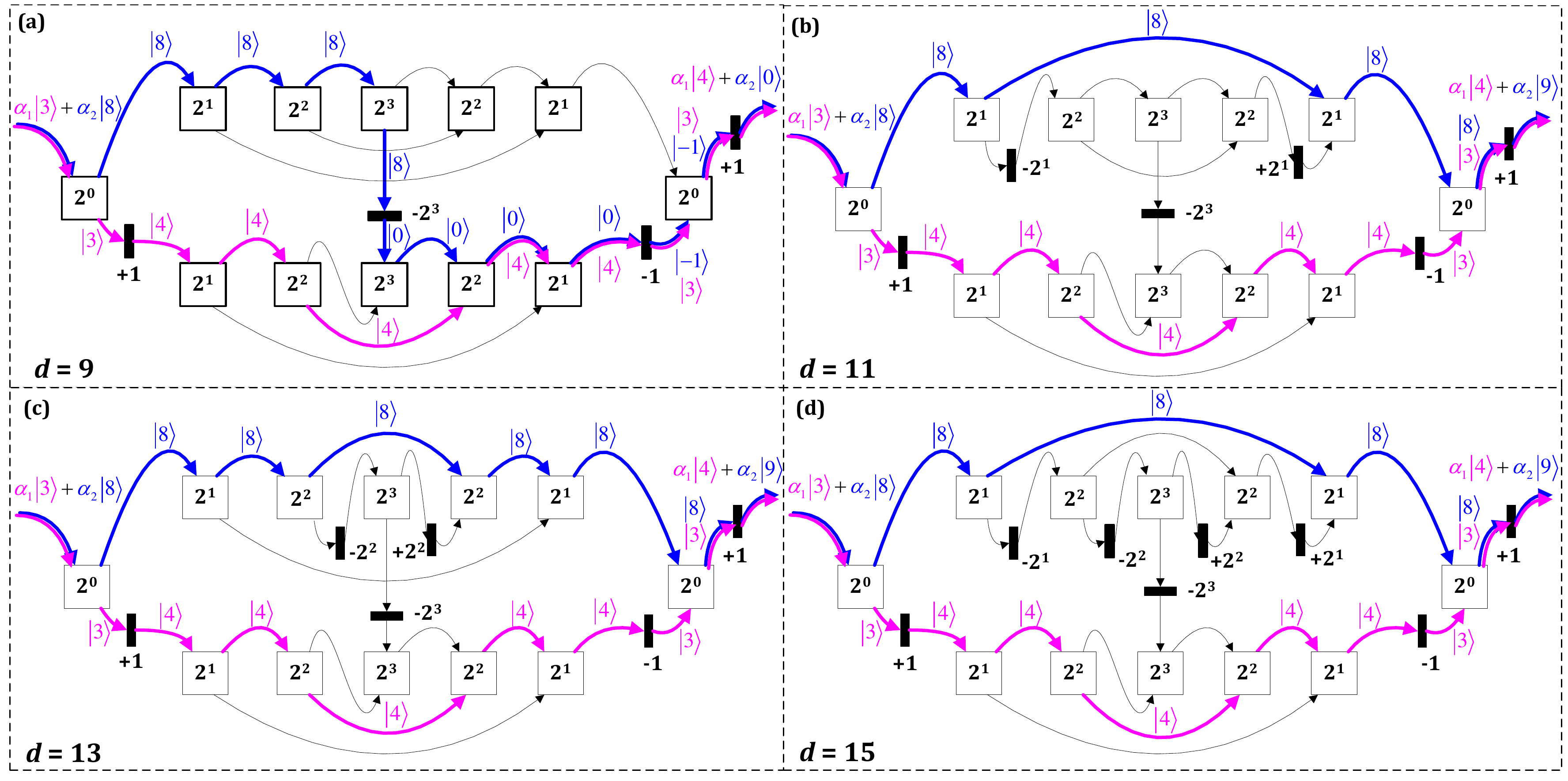}}
    \vspace{-0.3cm}%
    \caption{(a), (b), (c), and (d) The experimental setups for the $X$-gate in dimensions $d = 9, 11, 13$, and $15$. The number of OAM-BS is 12 in all cases. While individual elements are connected differently in each setup, the overall structure does not change. A superposition $\alpha_1 \ket{3} + \alpha_2  \ket{8}$ goes through different paths in different setups.}
\label{fig10:figure10}
\vspace{-0.3cm}%
  \end{minipage}\hfill
\end{figure*}

\clearpage

\subsection{Different basis states}

\begin{figure*}
  \centering
  \begin{minipage}[t]{0.96\textwidth}
    \centering
    \raisebox{-\height}{\includegraphics[width=1\textwidth]{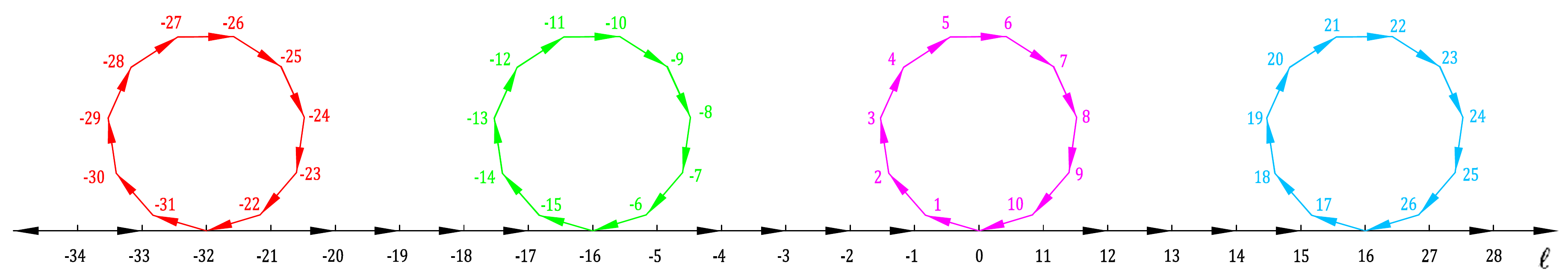}}
    \vspace{-0.4cm}%
    \caption{Possible 11-dimensional cyclic transformations implemented by the setup in figure \ref{fig5:figure5}(d). Each closed hendecagon shows the set of eleven OAM modes that can be cycled through with our experimental setup.}
\label{fig11:figure11}
\vspace{-0.1cm}%  
\end{minipage}\hfill
\end{figure*}

The experimental setup shown in figure \ref{fig5:figure5}(d)  implements the 11-dimensional cyclic transformation not only for states $\ket{0}, \ldots, \ket{10}$, but also for infinitely many additional sets of states, three of which are shown in figure \ref{fig11:figure11}. This feature also exists in other dimensions.

\subsection{Simplification}

The structure of setups generated by Algorithm \ref{alg:algorithm3} and demonstrated in figure \ref{fig5:figure5} is very symmetric --- in the first part individual modes entering the setup are rerouted into different paths, the second part treats the highest mode separately from the other modes, and finally the third part again recombines all modes into a single output path. Similarity of the third, recombination, part and the first, rerouting, part allows us to remove the third part altogether. Provided that the photon has a sufficiently large coherence length the original setup can therefore be simplified considerably. 

\begin{figure}[!htbp]
\includegraphics[width=0.48 \textwidth]{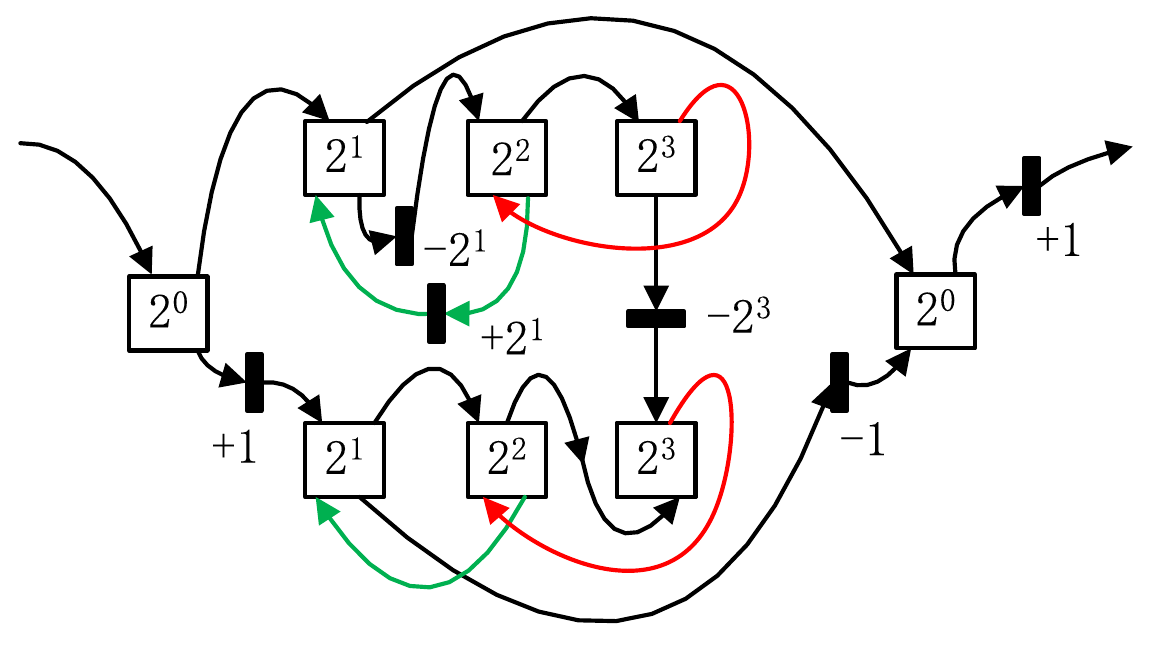}
\caption{The simplified experimental setup implementing the 11-dimensional cycle. In figure  \ref{fig5:figure5}(d), the right part is not necessary as the left part can be used both to reroute and to recombine individual OAM modes.}  
\centering
\label{fig7:figure7}
\end{figure}

As an example, we suppose that the photon has sufficiently long coherence length and simplify the 11-dimensional cycle in figure \ref{fig5:figure5}(d). Four OAM-BSs can thus be removed and paths leading to them are reconnected as demonstrated by red and green lines in figure \ref{fig7:figure7}. The number of OAM-BSs for dimension $d = 11$ is thus reduced from 12 to 8. For dimension 500, such a simplification leads to reduction from 28 to 16. In general, the total number of OAM-BSs used in the simplified setup is equal to
\begin{align}
N_{S} = M+2 \times \floors{\log_2 {Q}} + 2,
\label{eq:equation7}
\end{align}
which is approximately a half of the OAM beam-splitters used in the original setup (\emph{cf.} Eq. (\ref{eq:equation3})). Even though the scaling is still logarithmic, especially for high-dimensional cyclic transformations the reduction in complexity of actual experimental setups may be of great usefulness. 

\subsection{$X^{-1}$ gate}

\begin{figure}[!htbp]
\includegraphics[width=0.48 \textwidth]{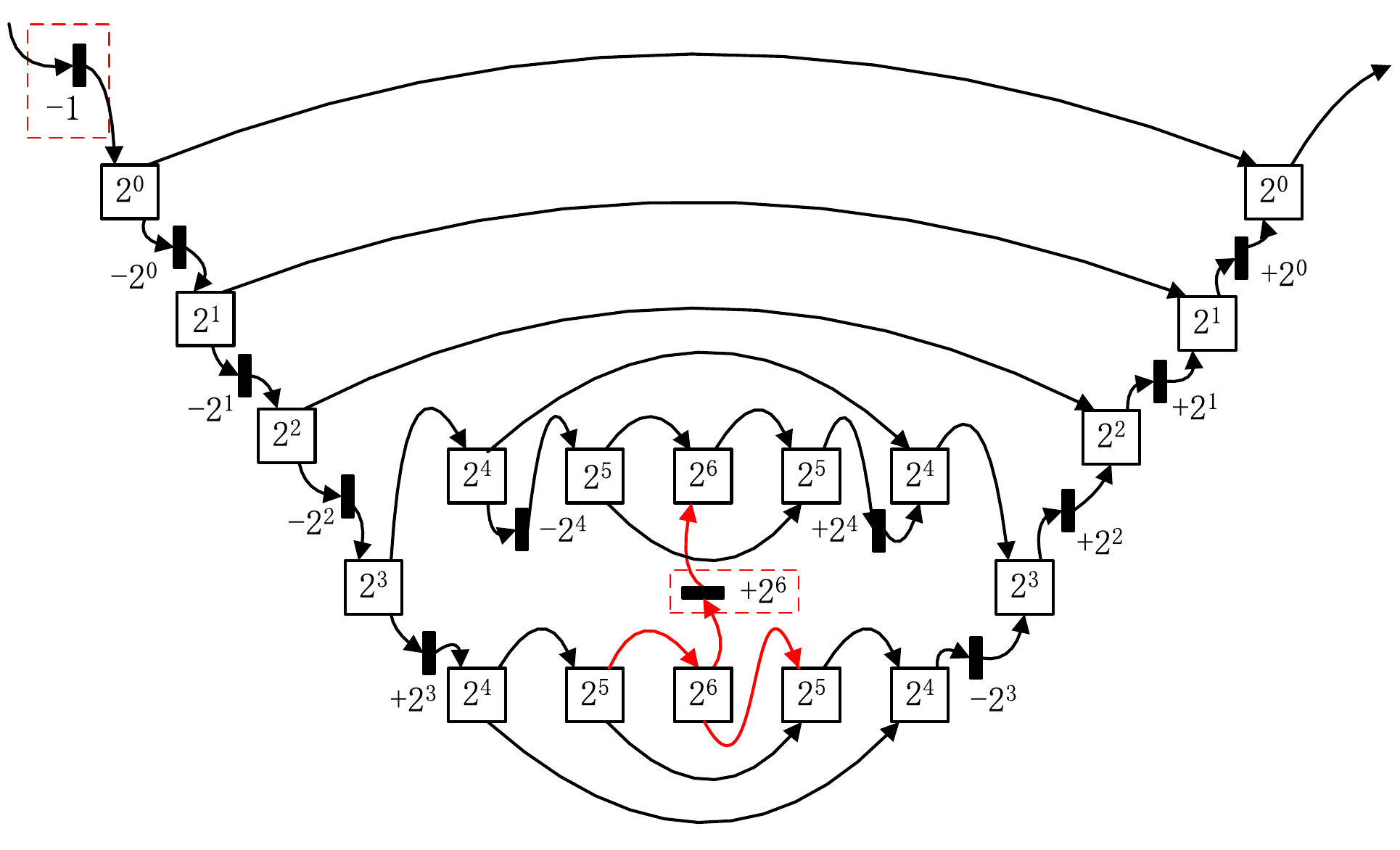}
\caption{The $X^{-1}$ gate implementation in the 88-dimensional state space. The differences from the implementation of the $X$-gate in figure \ref{fig5:figure5} are highlighted. Specifically, the final hologram is moved to the beginning, the central hologram is inverted and two central OAM-BSs are connected differently to their neighbors and to each other.}
\centering
\label{fig12:figure12}
\end{figure}

In figure \ref{fig12:figure12} one can see the setup for the $X^{-1}$ gate in the 88-dimensional Hilbert space, where differences from the implementation of the corresponding $X$-gate in figure \ref{fig5:figure5} are highlighted.

\end{document}